# IT AMBIDEXTERITY AND PATIENT AGILITY: THE MEDIATING ROLE OF DIGITAL DYNAMIC CAPABILITY

*Research Paper*

Van de Wetering, Rogier, Open University, the Netherlands, rogier.vandewetering@ou.nl

## Abstract

*Despite a wealth of attention for information technology (IT)-enabled transformation in healthcare research, limited attention has been given to IT's role in developing specific organizational capabilities to respond to patients' needs and wishes adequately. This paper investigates how hospital departments can leverage the equivocal capacity to explore and exploit IT resources and practices, i.e., IT ambidexterity, to adequately sense and respond to patients' needs and demands, i.e., patient agility. Following the dynamic capabilities view, this research develops a research model and tests it accordingly using data obtained from 107 clinical hospital departments from the Netherlands through an online survey. The model's hypothesized relationships are tested using structural equation modeling (SEM). The outcomes demonstrate the significance of IT ambidexterity in developing a digital dynamic capability that, in turn, positively influences patient agility. The study outcomes can be used to transform clinical practice and contribute to the current IS knowledge base.*

## 1 Introduction

Hospitals worldwide face a multitude of challenges to ensure high quality across the entire patient care delivery continuum. They are, e.g., burdened by exhaustive management, regulatory, and administrative processes. Under these turbulent conditions, hospitals can leverage new innovative information technologies (IT), such as electronic medical records, patient accessible electronic health records, artificial intelligence, the Internet of Things, and interoperable data and platforms, to enable the ubiquitous availability of patient information and enhance the quality of clinical practice. It, therefore, seems that hospitals are on the brink of a monumental digital transformation change.

Hospitals now have an opportunity to deploy new digital technologies to enhance their internal decision-making capability (Van de Wetering, 2018), redefine patient communication and engagement, and become true IT innovators (Leidner, Preston, & Chen, 2010). Medical professionals can also use innovative IT solutions and the exponential volumes of patient-generated data to enhance care delivery (McCullough, Casey, Moscovice, & Prasad, 2010). Notwithstanding, there are many cultural, social, technical, and organizational challenges in the process of fully leveraging digital technologies (Kohli & Tan, 2016; Kruse, Kristof, Jones, Mitchell, & Martinez, 2016; Sligo, Gauld, Roberts, & Villa, 2017). Moreover, the extant scholarship contends that IT could also hamper the process of gaining organizational benefits (Brynjolfsson & Hitt, 2000; Carr, 2003; Overby, Bharadwaj, & Sambamurthy, 2006). Thus, understanding the facets that drive IT investments benefits is very valuable in clinical settings (Hessels, Flynn, Cimiotti, Bakken, & Gershon, 2015).

Previous scholarship examined the role of IT and contributions to the so-called 'IT-enabled capabilities.' In particular, the role of IT human competencies and IT infrastructure capability have been explored as enablers of the formation of IT-enabled capabilities and organizational sensing and responding under turbulent conditions (Chakravarty, Grewal, & Sambamurthy, 2013; Fink, 2011; Rai & Tang, 2010; Roberts & Grover, 2012b; Van de Wetering, Versendaal, & Walraven, 2018). However, there are still crucial gaps that remain in the extant literature.

First, despite a wealth of attention for IT adoption and IT-enabled transformation in healthcare research (Andargoli, Scheepers, Rajendran, & Sohal, 2017; Chen, Yu, & Chen, 2015; Jones, Heaton, Rudin, & Schneider, 2012; Nair & Dreyfus, 2018; Yichuan Wang & Hajli, 2017; Wang, Kung, Wang, & Cegielski, 2018; Zhou & Piramuthu, 2018), limited attention has been given to the role of IT in developing specific organizational capabilities to respond to patient's needs and wishes adequately, and enhance patient engagement (Asagbra, Burke, & Liang, 2018; Bradley, Pratt, Thrasher, Byrd, & Thomas, 2012). Second, it currently remains unclear how hospital departments—that are responsible for patient care delivery—can utilize the equivocal capacity to 'explore' and 'exploit' IT resources and practices, i.e., IT ambidexterity (Lee, Sambamurthy, Lim, & Wei, 2015), to drive a hospitals' departments digital dynamic capability. Such a capability represents the degree to which qualities and competencies are present to manage innovative digital technologies for new exceptional and effective patient service development (Khin & Ho, 2019). Third, the extant literature unfolded critical insights into the competences and routines underlying dynamic capabilities—that represent an organization's ability 'to act' under changing circumstances (Cepeda & Vera, 2007; Winter, 2003)—and how they enhance the operational functioning of the firm (Protogerou, Caloghirou, & Lioukas, 2012; Van de Wetering, 2019; Wilden & Gudergan, 2015). However, there seems to be less consensus about IT resources' pivotal role in developing these dynamic capabilities (Li & Chan, 2019; Menachemi, Matthews, Ford, Hikmet, & Brooks, 2009). Fourth, scholars have predominantly explored the ambidexterity-benefits relationship on the organizational level (Jansen, Simsek, & Cao, 2012; O'Reilly III & Tushman, 2008). So, investigating the role of an intermediate (digital) capability construct in the value path at the hospital department (strategic business or competence unit) level is valuable as it has been seldom explored (Gerybadze, 1998; Lee et al., 2015; Pang, Lee, & DeLone, 2014).

Finally, previous studies focused on particular aspects of IT resources and assets (e.g., IT flexibility, system design, IT human and management capabilities, knowledge and data capabilities, and repositories) from an organizational agility perspective (Chen & Siau, 2012; Lu & Ramamurthy, 2011; Tallon & Pinsonneault, 2011). In doing so, these particular studies highlight the facilitating role of the synergistic IT exploration and exploitation processes but failed to conceptualize and test this particular capacity adequately. Unfolding the benefits of such a dual capacity to aim for two disparate things simultaneously using empirical data is very relevant. It is relevant from both a theoretical and practical perspective as ITs business value, and the preceding IT investments can be justified in a clinical setting (Hessels et al., 2015; Lee et al., 2015; Sabherwal & Jeyaraj, 2015; Schryen, 2013; Van de Wetering, 2018).

Against this background, and in alignment with the hospital industries' focus on clinical excellence and patient-centered care (Chiasson & Davidson, 2005; Liedtka, 1992), this paper acclaims that IT ambidexterity enhances the ability to sense and respond adequately to the patient's needs and demands, i.e., patient agility, by facilitating the intermediate digital capability-building process. In doing so, this study follows a practitioner-based approach (Devaraj & Kohli, 2003; McCracken, McIlwain, & Fottler, 2001). This study focuses on the department level and patient agility, can, therefore, be considered the degree to which a department can sense and respond quickly to patient-based opportunities for innovation and competitive action.

This work is likewise relevant for clinical practice, as hospitals are currently exploring their digital options and digitally transforming their clinical processes using, e.g., mobile handheld devices and apps to increase error prevention, improve patient-centered care and provide ways for clinicians to be more agile in their work (Bradley et al., 2012; Devaraj, Ow, & Kohli, 2013; Mosa, Yoo, & Sheets, 2012; Sim, 2019). Also, hospitals in the Netherlands are bound to specific turnover ceilings agreements between hospitals and health insurers. The Dutch Healthcare Authority (NZa), an autonomous administrative authority falling under the Dutch Ministry of Health, Welfare, and Sport, ensures that these formal agreements focus specifically on patients' value rather than production volumes. Therefore, achieving patient agility in clinical practice is very valuable.

Hence, this research attempts to address the following research questions:

*(1) What is IT ambidexterity's effect on the hospital departments' patient agility and, thus, its ability to sense and respond timely and adequately to the patient's needs and demands? Furthermore,*

*(2) What is the particular role of digital dynamic capability in converting IT ambidexterity's effect to the hospital departments' patient agility?*

By addressing these crucial questions, this paper contributes to the medical informatics and information systems (IS) literature by unfolding the mechanisms through which the dual capacity of IT exploration and IT exploitation simultaneously drives patient agility in hospital departments. In addressing these questions, this study embraces the dynamic capabilities view (DCV) to employ a strong academic foundation with accompanying validated measurements (Khin & Ho, 2019; Lee et al., 2015; Roberts & Grover, 2012b; Van de Wetering, 2020).

This paper proceeds as follows. First, it reviews the theoretical development by highlighting key literature on IT resources and ambidexterity, the dynamic capabilities view, and organizational agility. Then, section 3 highlights the study's research model and associated hypotheses. Section 4 details the methods used in this study, after which section 5 outlines the results. This study ends by discussing the outcomes, including theoretical and practical contributions, and ends with concluding remarks.

## 2 Theoretical development

### 2.1 IT resources and IT ambidexterity

Scholars and practitioners started to build upon strategic management theories in the late '80 and early '90 and argue that organizations can sustain a competitive advantage due to their IT resources they own or have under their control (Wade & Hulland, 2004). However, the literature documents that IT investments do not always yield the presumed and anticipated results and may even impede the process of rapidly adjusting to the competitive environment (Brynjolfsson & Hitt, 1998; Carr, 2003; Overby et al., 2006; Strassmann, 1990). The central claim within the "resource-based" studies, in the context of IT, is that the adoption, deployment, and practices of IT as a unique and difficult-to-imitate resource, creates business value (Bharadwaj, 2000; Wade & Hulland, 2004; Wang, Liang, Zhong, Xue, & Xiao, 2012).

Previous studies contended that IT resources are crucial in the formation of technological-driven capabilities (Aral & Weill, 2007; Ross, Beath, & Goodhue, 1996; Van de Wetering et al., 2018; Weill, Subramani, & Broadbent, 2002), and therefore, considered a key strategic priority for organizations (Bharadwaj, 2000), even so in healthcare (Bardhan & Thouin, 2013; Wang, Kung, & Byrd, 2018). However, obtaining the value from IT resources is not a straightforward process. Instead, the extant literature contends that the business value results from leveraging and aligning complementary IT resources (Pavlou & El Sawy, 2006; Sheikh, Sood, & Bates, 2015; Van de Wetering, 2018; Wade & Hulland, 2004). This process also applies to healthcare, where senior executives and hospital management want to leverage their IT and data resource investment successfully (Van de Wetering, 2018; Wang, Kung, & Byrd, 2018).

In practice, organizations need to pursue and deal with two seemingly opposing modes of operandi, i.e., the ability to adapt existing IT resources to the current business environment and demands and focus on developing IT resources that contribute to long-term organizational benefits (March, 1991). The literature refers to this phenomenon as 'ambidexterity' (Gibson & Birkinshaw, 2004; Jansen, Van Den Bosch, & Volberda, 2006; Raisch, Birkinshaw, Probst, & Tushman, 2009; Tushman & O'Reilly III, 1996). The simultaneous alignment of 'exploration' and 'exploitation' by organizations will likely provide them with sustained competitive benefits (Gibson & Birkinshaw, 2004; Jansen et al., 2006; Junni, Sarala, Taras, & Tarba, 2013; Raisch et al., 2009). The notion of IT ambidexterity, as defined by Lee et al. (2015, p. 398) as *"..the ability of firms to simultaneously explore new IT resources and* *practices (IT exploration) as well as exploit their current IT resources and practices (IT exploitation)*" is a fundamental capability that builds upon the ambidexterity and IT resources and capability-building perspective. Ambidexterity gained serious attention of over the past few years (Chang, Wong, Eze, & Lee, 2019; Syed, Blome, & Papadopoulos, 2020). IT exploration concerns the organization's efforts to pursue new knowledge and IT resources (Lee et al., 2015; March, 1991). On the other hand, IT exploitation is typically conceptualized as a construct that captures the degree to which organizations take advantage of existing IT resources and assets, e.g., reusing existing IT applications and services for new patient services and the reuse of existing IT skills.

## 2.2 Dynamic capabilities view

The dynamic capabilities view (DCV) is a leading theoretical framework in the field of strategic management and information systems (Di Stefano, Peteraf, & Verona, 2014; Pavlou & El Sawy, 2011; Schilke, 2014). Under conditions of high economic turbulence and uncertainty, the theory argues that traditional resource-based capabilities do not provide organizations with a competitive edge (Drnevich & Kriauciunas, 2011; Teece, Peteraf, & Leih, 2016; Wilden & Gudergan, 2015). Instead, it is firms' ability to integrate, build, and reconfigure internal and external competences to capitalize on rapidly changing environments that explain how organizations can obtain and maintain a competitive edge (Teece, Pisano, & Shuen, 1997).

The DCV highlight two crucial aspects that were not the primary focus under resource-based approaches (Wernerfelt, 1984), i.e., 'dynamic' and 'capabilities.' Dynamic refers to firms' capacity to renew competences and capacities to obtain congruence with rapidly changing business environment (Teece et al., 1997). Capabilities, on the other hand, focus on purposefully adapting the firms' resource base. According to Teece (Teece et al., 1997, p. 515), these capabilities emphasize "…*the key role of strategic management in appropriately adapting, integrating, and reconfiguring internal and external organizational skills resources, and functional competences to match the requirements of a changing environment.*" Hence, the DCV can be considered an integrative approach to understanding the newer sources of competitive advantage and firm's ability 'to act' under changing circumstances (Cepeda & Vera, 2007; Winter, 2003) and has been widely validated through empirical studies also in healthcare (Pablo, Reay, Dewald, & Casebeer, 2007; Singh, Mathiassen, Stachura, & Astapova, 2011; Wu & Hu, 2012).

### 2.2.1 The concept of organizational and patient agility

Organizational agility is considered to be a manifested type of dynamic capability (Teece et al., 2016) and is especially influential among agility studies published in the Basket of Eight, see for instance, (Lee et al., 2015; Queiroz, Tallon, Sharma, & Coltman, 2018; Sambamurthy, Bharadwaj, & Grover, 2003). It can be conceptualized as a dynamic capability if "*they permit organizations to repurpose or reposition their resources as conditions shift.*" (Tallon et al., 2019, p. 220).

As such, the concept of organizational agility has been proposed under the DCV as a crucial organizational capability to respond to changing conditions while simultaneously proactively enacting on the dynamic environment regarding customer demands, supply chains, new technologies, governmental regulations, and competition (Park, El Sawy, & Fiss, 2017; Roberts & Grover, 2012b; Teece et al., 2016). The environment imposes several contingencies (Lu & Ramamurthy, 2011; Park et al., 2017). Therefore, Overby (2006) and Ravichandran (2018) argue it is vital for firms to adapt, adjust, and renew their working systems and procedures to enhance their rent yielding potential and seize market opportunities. This 'sense-and-respond' capability has been defined and conceptualized in many ways and through various theoretical lenses in the IS literature (Chakravarty et al., 2013; Sambamurthy et al., 2003; Vickery, Droge, Setia, & Sambamurthy, 2010). Several scholars argue that, even though there are ambiguities in the definitions that are likewise reflected in operationalized conceptualizations, several high-level characteristics can be devised from the extant literature that view organizational agility as a higher-order multidimensional construct (Lee et al., 2015; Overby et al., 2006; Roberts & Grover, 2012a). From the literature, two high-level organizational routines can be synthesized: deliberately 'sensing' and 'responding' to business events in the process of capturing business and market opportunities. These two capabilities are essential for an organization's competitive advantage (Overby et al., 2006).

There are no studies (empirical nor theoretical) that conceptualize hospital departments' patient agility, following the dynamic capabilities view. It is, therefore that this current article perceives patient agility as a higher-order manifested type of dynamic capability that allows hospital departments to adequately 'sense' and 'respond' to patient-based opportunities, needs, demands within a fast-paced hospital ecosystem context (Roberts & Grover, 2012a; Teece et al., 2016).

### 2.2.2 Digital dynamic capability

The concept of digital dynamic capability builds upon a rich foundation of the previously discussed DCV. According to (Khin & Ho, 2019, p. 4), digital dynamic capabilities can be considered the *"organization's skill, talent, and expertise to manage digital technologies for new product development."* This capability is essential for a hospital to master digital technologies, drive digital transformations, and develop innovative patient-centered services and products. For this study, a hierarchical capability view is adopted following previous scholarship (Božič & Dimovski, 2019; Danneels, 2002; Winter, 2003). As such, the digital dynamic capability is conceptualized as a lower-order technical dynamic capability that facilitates the process of developing higher-order dynamic organizational capabilities such as innovation ambidexterity, absorptive capacity, and organizational adaptiveness (Božič & Dimovski, 2019; Li & Chan, 2019; Wang & Ahmed, 2007).

This conceptualization is also in line with the previous scholarship that conceptualized technological capabilities as a technical dynamic capability. For instance, Benitez et al. (2018) conceptualized a flexible IT infrastructure as a dynamic capability that provides the organization with adequate responsiveness by enabling the business flexibility to sense and seize mergers and acquisition opportunities (Benitez, Ray, & Henseler, 2018). Likewise, Queiroz et al. (2018) argue that IT application orchestration, i.e., the ability to renew IT application portfolios, should be conceived as a dynamic capability and enhances competitive firm performance. In practice, this capability requires specific, idiosyncratic, and heterogeneous competencies to develop and is, therefore tough to mimic and establish within organizations (Božič & Dimovski, 2019; Khin & Ho, 2019; Teece et al., 1997; Tripsas, 1997).

## 3 Research model and hypotheses

Following the theoretical development section, IT ambidexterity as a core organizational IT resource is expected to enhance hospital departments' level of patient sensing and responding capability (both conceptualized as higher-order dynamic capability) through digital dynamic capability (as a lower-order technical dynamic capability). Figure 1 demonstrates the research model and the associated hypotheses that will be clarified below.

A firm's deliberate IT investments are crucial for the process of capability-building and gaining IT business value under turbulent conditions (Overby et al., 2006; Sambamurthy et al., 2003; Setia, Venkatesh, & Joglekar, 2013). In this discussion, IT resources are typically referred to by their aggregated latent components and qualities (e.g., hardware, software, networks, data sources) and IT-related managerial activities (e.g., IT planning, business connectivity) and how they related to business value (Kim, Shin, Kim, & Lee, 2011; Tippins & Sohi, 2003). However, recent studies argue that IT-business value and organizational agility do not result from the deployment of isolated (non)IT resources and competencies. Instead, IT-business value seems to emerge from the complementarity to assimilate and re-align the IT resource portfolio to the changing business needs and demands (Ravichandran, 2018; Walraven, Van de Wetering, Versendaal, & Caniëls, 2019).

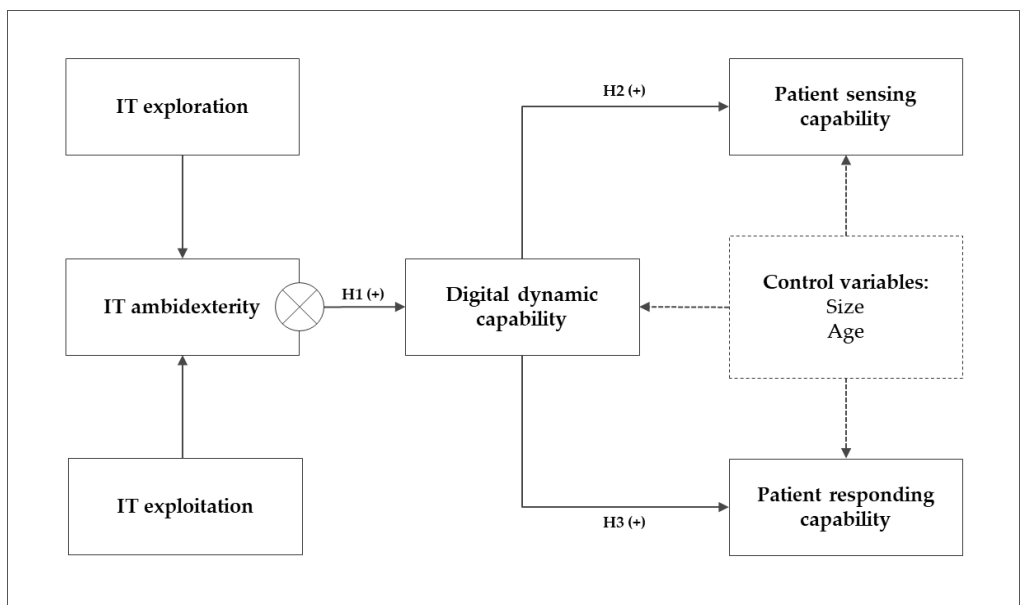

*Figure 1. Research model*

Hospitals need to deal with many challenges (e.g., organizational, social, cultural) (Bhattacherjee & Hikmet, 2007; Chaudhry et al., 2006; Kohli & Tan, 2016). Therefore, they must embrace an ambidextrous IT implementation strategy so that short-term exploitation of (existing) IT resources is balanced with an exploratory mode that drives IT-driven business transformation (Gregory, Keil, Muntermann, & Mähring, 2015). It is only when these two modes are in sync that hospital departments' are better-equipped to develop digital capabilities and frame the hospital's business strategy and clinical practice (Jaana, Ward, & Bahensky, 2012; Khin & Ho, 2019; Van de Wetering et al., 2018).

IT exploration facilitates the experimentation and usage of new IT resources (e.g., new collaborative research platforms, decision-support systems, big data, and clinical analytics, social media) so that upon success can serve as a basis to reshape the current patient engagement and care. On the other hand, IT exploitation is instead focused on using, enhancing, and repackaging existing IT resources (e.g., reuse or redesigning current EMR for new patient service development and ensuring hospital-wide accessibility to clinical patient data and information). IT exploitation allows departments to reuse existing modular and compatible IT-infrastructures and software components and integrate them with their daily business operations and clinical practices (Lee et al., 2015; Tarenskeen, Van de Wetering, Bakker, & Brinkkemper, 2020).

It is expected that the simultaneous engagement of IT exploration and IT exploitation will strengthen hospital departments' digital options and competencies to manage innovative digital technologies for new patient service as these two distinct IT ambidexterity qualities in isolation are not likely to enhance the hospital department's digital dynamic capabilities (Sambamurthy et al., 2003; Van de Wetering et al., 2018; Voudouris, Lioukas, Iatrelli, & Caloghirou, 2012). In line with this reasoning, this study now contends that the complementary effect underlying IT ambidexterity facilitates hospital departments to develop a digital dynamic capability that serves as a basis to enhance patient agility. Moreover, such a strategy enables hospital departments to alleviate possible (unforeseen) risks associated with exploratory and exploitatory modes. Hence, the following hypothesis is defined:

**Hypothesis 1:** *IT ambidexterity will positively enhance hospital departments' level of digital dynamic capability.*

Digital dynamic capability is a crucial dynamic capability necessary to innovate and enhance business operations (Acur, Kandemir, De Weerd-Nederhof, & Song, 2010; Khin & Ho, 2019; Li & Chan, 2019;

Zhou & Wu, 2010). Various prior studies investigated the benefits that result from developing a digital dynamic capability. Wang et al. (2004), for instance, argue that digital dynamic capability allows leveraging IT and knowledge resources to deliver innovative services that customers value and contribute to organizational benefits. Coombs and Bierly (Coombs & Bierly III, 2006) empirically showed that a sophisticated digital dynamic capability enables competitive advantages. Thus, the extant literature shows that digital dynamic capability drives organizations' ability to learn from experience in turbulent economic and competitive environments. Hence, in such an environment, it is essential to search continuously, identify, and absorb new technological innovation such that they can be used to respond to changing customer behavior, demands, and wishes timely, adequately, and innovatively (Acur et al., 2010; Roberts & Grover, 2012b). This outcome is likewise consistent with results from Westerman et al. (2012), Khin et al. (2019), and Ritter and Pedersen (2019), who showed that digital dynamic capability is crucial to deploy new innovative business models, enhance customer experiences, and improve business operations. By actively managing the opportunities provided by innovative technologies and responding to digital transformation, organizations can succeed in their digital options and services (Khin & Ho, 2019).

A technological-driven capability is crucial for hospital departments that want to strive for patient agility in clinical practice because the process of achieving new digital patient service solutions is exceedingly dependent on its ability to manage digital technologies (Khin & Ho, 2019). It requires proactively responding to digital transformation, mastering the state-of-the-art digital technologies, and deliberately developing innovative patient services using digital technology. Such a capability goes well beyond the notion of IT capabilities, i.e., aggregation of IT resources and IT competencies in the vast majority of empirical studies (Chen et al., 2014; Kim et al., 2011; Wade & Hulland, 2004).

The digital dynamic capability allows hospital departments, e.g., to absorb and process sensitive patient information better, support clinicians in their decision-making processes, exchange clinical data, and facilitate patient health data accessibility (Van de Wetering et al., 2018). Hence, hospitals that actively invest and develop such a capability are likely to anticipate their patients' needs (of which they might be physically and mentally unaware) and respond fast to changes in the patient's health service needs using digital innovations and assessments of clinical outcomes (Khin & Ho, 2019; Roberts & Grover, 2012a). Therefore, such a strategically significant capability is crucial for the departments' focus on quality, efficiency, and enhancing the patient's clinical journey. Based on the argument given above and building upon the DCV, the following two hypotheses are defined:

**Hypothesis 2:** *Hospital departments' digital dynamic capability will be positively associated with a patient sensing capability.*

**Hypothesis 3:** *Hospital departments' digital dynamic capability will be positively associated with a patient responding capability.*

# 4    Methods

## 4.1    Data collection

Survey data were systematically collected using an online survey that contained all questions to test the study's model and hypothesized relationships. This survey was pretested on several occasions by five Master students (that do their Master's thesis research) and six medical practitioners and scholars to improve both the content and face validity of the survey items. These respondents all had sufficient knowledge and experience to assess the survey items effectively to provide valuable improvement suggestions. The data were finally cross-sectionally collected from University medical centers, top clinical training hospitals, and general hospitals. The target population was (clinical) department heads, team-leads, managers, and doctors in line with the study objectives. These respondents are the foremost respondents—at the hospital department level—who can provide insights into the unique and sometimes complicated situations where medical knowledge is exploited, enabling a unique treatment course (Wu & Hu, 2012). Moreover, they actively contact patients or have an intelligible insight into the department's patient interactions, and IT use.

Data were conveniently sampled from Dutch hospitals through the 5 Master students' professional networks within hospitals using email, telephone, and social networks. The final data collection took place between November 10[th,] 2019, to January 5[th,] 2020. Also, anonymity for the respondents was guaranteed. The online survey tool registered 230 active and unique respondents. In total, 101 cases were removed from the data because of unreliable data entries or no entries. Finally, 21 additional respondents were removed due to substantial missing values, and one respondent was removed as the function did not belong to our target population. This study uses 107 complete survey responses for final analyses. This study thus uses a single informant to fill in the survey for the entire department. Within the obtained sample, 36 respondents work for a University medical center (33.6%), 41 work for a specialized top clinical (training) hospital (38.3%), and the final 30 work for general hospitals (28%). Table 1 shows the demographics of the obtained data sample.

| Element | Category | Frequency | Percentage |
|---|---|---|---|
| Hospital type | University medical center | 36 | 33.6% |
| | Top clinical training hospital | 41 | 38.3% |
| | General Hospital | 30 | 28% |
| Department age | 0–5 years | 28 | 26,2% |
| | 6–10 years | 20 | 18,7% |
| | 11–20 years | 20 | 18,7% |
| | 20–25 years | 8 | 7,5% |
| | Over 25 years | 31 | 29,0% |
| Experience at this particular department | 0–5 years | 49 | 45,8% |
| | 6–10 years | 18 | 16,8% |
| | 11–20 years | 28 | 26,2% |
| | 20–25 years | 6 | 5,6% |
| | Over 25 years | 6 | 5,6% |
| Amount of patients | < 4000 | 25 | 23,4% |
| | 4000 – 6500 | 21 | 19,6% |
| | 6500 – 9000 | 12 | 11,2% |
| | 9000 – 11500 | 12 | 11,2% |
| | 11500 – 14000 | 11 | 10,3% |
| | > 14000 | 26 | 24,3% |

*Table 1. Demographics of participating hospital departments*

Most survey respondents are medical heads/chairs of the department (55.1%)[1], 27.1% is a practicing doctor, 12.1% is department manager, while the remaining 5.6% of the respondents hold other positions such as specialized oncology nurses. The average number of doctors for all the participating departments is 29.8, while the average total number of employees (including support staff) is 149.9.

This study accounts for possible non-response bias by using a T-test to assess whether or not there is a substantial (and significant) difference between the early respondents (*N*=66) and the late subsample (*N*=41 respondents) on the responses on the Likert scale questions. No significant difference could be detected. Finally, Harman's single-factor analysis was applied using exploratory factor analysis (in using IBM SPSS Statistics™ v24) to restrain possible common method bias (Podsakoff, MacKenzie, Lee, & Podsakoff, 2003; Richardson, Simmering, & Sturman, 2009). Hence, the current study sample is not affected by method biases, as no single factor attributed to the majority of the variance.

---

[1] 5 respondents claimed that they were team leads.

## 4.2 Measures and items

The selection of indicators was made based on previous empirical and validated work to increase the questions' internal validity and reliability. Since this research was done in a healthcare setting, some of the original items have to be re-worded to suit the context of Dutch healthcare.

This study operationalized *IT ambidexterity* using the item-level interaction terms of IT exploration and IT exploitation before running the structural model (Gibson & Birkinshaw, 2004; Lee et al., 2015). Measures for IT ambidexterity were adopted from (Lee et al., 2015). This study devised three core items from Khin and Ho (2019) to measure *digital dynamic capability* and conceptualized *patient agility* as a higher-order dynamic capability comprising the dimensions *'patient sensing capability'* and *'patient responding capability'* (Roberts & Grover, 2012b; Sambamurthy et al., 2003). This study adopts five measures for each of these two capabilities based on Roberts and Grover (2012b).

The constructs' items in the research model used a seven-point Likert scale ranging from 1: strongly disagree to 7: strongly agree, which is a commonly used classification in empirical survey studies since no archival data exist for quantifying the incorporated competencies and capabilities (Kumar, Stern, & Anderson, 1993). Following prior IS and management studies, we controlled for 'size' (full-time employees), operationalizing this measure using the natural log (i.e., log-normally distributed) and 'age' of the department (5-point Likert scale 1: 0–5years; 5: 25+ years). Table 2 includes all the constructs' items.

## 4.3 Model estimation and sample justification

The research model is assessed using a Partial Least Squares (PLS) structural equation modeling (SEM) application, i.e., SmartPLS version 3.2.9. (Ringle, Wende, & Becker, 2015), to estimate and run model parameter estimates. PLS allows researchers to confirm that measurement model—that runs estimates of the included latent constructs as weighted sums of a specific subset of the associated manifest variables —while also assessing the model's structural model that enables the process of testing hypothesized relationships (Hair Jr, Hult, Ringle, & Sarstedt, 2016). A pivotal reason to justify applying a variance-based approach to SEM is that it is appropriate in exploratory contexts and for the objective of theory development (Hair Jr, Sarstedt, Ringle, & Gudergan, 2017). PLS-SEM emphasizes prediction-oriented work (as is the case in this research) as to which PLS maximizes the proportion of explained variance ($R^2$) for all dependent constructs in the research model (Hair Jr et al., 2017). The estimation procedure uses the general recommended path weighting scheme algorithm (Ringle et al., 2015).

The current sample size of 107 is relatively small but exceeds the minimum threshold values to obtain stable PLS outcomes (Hair, Ringle, & Sarstedt, 2011). To further substantiate this claim, a power analysis was done using G*Power (Faul, Erdfelder, Buchner, & Lang, 2009). This study applies the commonly used statistical power level of 80%, an effect size of 0.15, and a 5% probability of error as input parameters. The maximum number of predictors in the research model is two (when including the non-hypothesized direct effect of IT ambidexterity on sensing and responding capability). G* Power's output parameters show that a minimum sample of 68 cases were needed, which is far below the current sample size of 107.

# 5 Empirical results

## 5.1 Reliability and validity assessments

As part of the measurement model assessment, this study evaluated the internal consistency reliability—using the composite reliability estimation (CR)—and convergent validity—using the average variance extracted (AVE)—of the constructs (Ringle et al., 2015). Also, all the construct-to-item loadings were investigated. All CR outcomes are well beyond 0.85, showing that the reflective constructs measure the same phenomenon, and the AVE-values greatly exceed the minimum threshold of 0.5. Discriminant validity is assessed using cross-loadings, the AVE's square root, i.e., the Fornell-Larcker criterion, and the newly developed heterotrait-monotrait ratio (HTMT) criterion as a metric of proper correlations among the model's constructs (Henseler, Ringle, & Sarstedt, 2015).

| Constructs/items | |
|---|---|
| **IT ambidexterity** (Lee et al., 2015) | **Patient agility** (Roberts & Grover, 2012a) |
| *IT exploration capability* | *Patient sensing capability* |
| Acquire new IT resources (e.g., potential IT applications, critical IT skills) | We continuously discover additional needs of our patients of which they are unaware |
| Experiment with new IT resources | We extrapolate key trends for insights on what patients will need in the future |
| Experiment with new IT management practices | We continuously anticipate our patients' needs even before they are aware of them |
| *IT exploitation capability* | We attempt to develop new ways of looking at patients and their needs |
| Reuse existing IT components, such as hardware and network resources | We sense our patient's needs even before they are aware of them |
| Reuse existing IT applications and services | *Patient responding capability* |
| Reuse existing IT skills | We respond rapidly if something important happens with regard to our patients |
| **Digital dynamic capability** (Khin & Ho, 2019) | We quickly implement our planned activities with regard to patients |
| Responding to digital transformation | We quickly react to fundamental changes with regard to our patients |
| Mastering the state-of-the-art digital technologies | When we identify a new patient need, we are quick to respond to it |
| Developing innovative patient services using digital technology | We are fast to respond to changes in our patient's health service needs |

*Table 2.        Construct and items.*

Analyses of the cross-loading show that each measurement item loads more strongly on its associated construct than other constructs (Farrell, 2010). Also, the Fornell-Larcker criterion for each of the constructs is higher than the inter-construct correlations. Finally, the HTMT assessment results indicate that each correlation score is well below the conservative $HTMT_{0.85}$ mark (Henseler et al., 2015). These outcomes confirm the model's constructs' discriminant validity.

## 5.2 Structural model assessment

A non-parametric bootstrapping approach was applied in SmartPLS to obtain the (regression) coefficients' significance levels among the constructs in the structural model. Hence, 5000 subsampling bootstraps were used with observations randomly drawn from the original set of data (i.e., the sample of 107) (Hair Jr et al., 2016). Also, this study investigated the strength of the coefficients of determination ($R^2$), the effect sizes ($f^2$), and the model's predictive power calculated using Stone-Geisser $Q^2$ values (Hair Jr et al., 2016). The established structural paths results show that IT ambidexterity positively influences digital dynamic capability ($β = 0.689; p < 0.0001$). The explained variance value suggests that 47.4% of the digital dynamic capability variance can be explained by IT ambidexterity. This degree of predictive accuracy can be classified as moderate to substantial (Chin, 1998).

| | Assessed path | Effect / $f^2$ | *t*-value | Conclusion |
|---|---|---|---|---|
| **H1** | IT ambidexterity → digital dynamic capability | 0.689 / 0.903 | 13.692 | Significant |
| **H2** | digital dynamic capability → patient sensing capability | 0.593 / 0.542 | 8.735 | Significant |
| **H3** | digital dynamic capability → patient responding capability | 0.463 / 0.273 | 5.636 | Significant[1] |
| *Additional path ($P_i$) and control estimations ($C_i$) using bootstrapping* | | | | |
| *P1* | IT ambidexterity → patient sensing capability | 0.176 / 0.026 | 1.442 | Not significant[2] |
| *P2* | IT ambidexterity → patient responding capability | 0.095 / 0.006 | 0.809 | Not significant[2] |
| *C1*[4] | Size → patient sensing capability | 0.004 / 0.000 | 0.048 | No confounding |
| *C2* | Size → patient responding capability | -0.048 / 0.003 | 0.536 | No confounding |
| *C3* | Size → digital dynamic capability | -0.056 / 0.006 | 0.780 | No confounding |
| *C4* | Age → patient sensing capability | -0.066 / 0.007 | 0.965 | No confounding |
| *C5* | Age → patient responding capability | -0.043 / 0.002 | 0.495 | No confounding |
| *C6* | Age → digital dynamic capability | -0.008 / 0.000 | 0.119 | No confounding |
| [1] *The explained variance ($R^2$) is comparably lower (60%) than the $R^2$ for patient sensing capability.* <br> [2] *Digital dynamic capability fully mediates IT ambidexterity on patient sensing and responding capability* | | | | |

*Table 3. Structural path results.*

Likewise, support was found for both the following paths, i.e., digital dynamic capability → patient sensing capability ($β = 0.593; p < 0.0001$), and digital capability → patient responding capability ($β = 0.463; p < 0.0001$). The structural model outcomes show that the total explained variance for patient sensing capability after removing non-significant control variables is 35.1% ($R^2 = 0.351$); this amount also exhibits a moderate effect. The $R^2$ for patient responding capability is 21.4%. This accuracy level is still deemed adequate but less strong than the explained variance for patient sensing capability (Chin, 1998).

Both 'age' and size' (i.e., the control variables) showed non-significant effects expelling possible confounding issues. The $Q^2$ values obtained through the blindfolding show that the $Q^2$ value (for digital dynamic capability) is well above zero ($Q^2 = 0.353$). Also, patient sensing capability and patient responding capability respectively showcase high $Q^2$ values ($Q^2 = 0.242; Q^2 = 0.157$). These outcomes likewise confirm the model's predictive relevancy (Hair Jr et al., 2016).

Table 3 summarizes the structural model analyses' outcomes, including the effect sizes, specific bootstrap t-values, and additional path analyses.

# 6 Discussion, implications, and concluding remarks

The digital transformation brings about an unprecedented challenge for modern-day hospitals (Agarwal, Gao, DesRoches, & Jha, 2010). Decision-makers and stakeholders across the hospital need to make sure that digital innovations are aligned and deployed with care so that they enhance efficiencies, decision-making, quality of services, so that personalized and patient-centered care can be delivered (McGrail, Ahuja, & Leaver, 2017; Van de Wetering, 2018; Walraven et al., 2019). From a research perspective, there is still a limited understanding of how IT resources and the digital capability-building processes can facilitate patient agility and contribute to the much-needed insights on obtaining value from IT at the departmental level. This study aimed at addressing these particular gaps in the literature.

## 6.1 Contributions to theory

This study designed and tested a research model that argues that the hospital department's capacity to explore and exploit IT resources and practices equivocally would drive a department's patient agility by first enabling digital dynamic capability. Outcomes of this study found support for this claim. The structural model analyses unfolded that a hospital department's ability to simultaneously pursue 'exploration' and 'exploitation' in their management of IT practices is a crucial driver of digital dynamic capability, and thus, a necessity to integrate digital technologies with the digital talent of doctors and medical practitioners and to be responsive in the process of patient care delivery (Khin & Ho, 2019). This study shows that digital dynamic capability, in turn, enhances the conceptualized construct of patient agility by sequentially enhancing patient sensing capability and patient responding capability. These results collectively address the research questions; the effect of IT ambidexterity on patient agility is indirect and fully mediated by digital dynamic capability.

These results are coherent with previous work prompting that those hospital departments that invest and enhance their skills, competences, and knowledge in managing innovative digital technologies are better equipped to be responsive, innovative, and satisfy patients' needs (Bolívar-Ramos, García-Morales, & García-Sánchez, 2012; Khin & Ho, 2019; Singh et al., 2011; Wu & Hu, 2012).

Notwithstanding, these current outcomes add to the current growing body of knowledge on the degree to which IT resources and competencies contribute to organizational capabilities and benefits (Chakravarty et al., 2013; Chen et al., 2014; Lu & Ramamurthy, 2011). Specifically, this study provides insights into the much-needed intermediate (and mediating) role of digital dynamic capability (Lee et al., 2015) and at the scarcely examined capability developing process at the departmental level (Gibson & Birkinshaw, 2004; Jansen et al., 2012). This study also advances the current insights on the resource and capability-building perspective (Lu & Ramamurthy, 2011; Pang et al., 2014; Sambamurthy et al., 2003; Teece et al., 1997) by unfolding the nomological path from 'resources' to the 'IT-enabled value-perspective.' It does so by showing that hospitals—that are committed to the process of ambidextrously managing their IT resources—are more proficient in promptly sensing and responding to patients' medical needs and demands. These theoretical contributions are valuable as these particular insights remained unclear in the extant literature, and future research can take these particular insights into account when investigating the IT benefits in hospitals.

## 6.2 Practical implications

The study also offers various practical implications for hospital department managers and practitioners. First, this study unfolds the critical resources and capabilities that hospital departments can leverage from a patient agility perspective. In doing so, this work embraces a dynamic capabilities view when it comes to IT resources deployments. Hence, hospital enterprises must direct IT investments to bring about the highest degree of IT business value, given the many substantial challenges to ensure high quality across the patient care delivery continuum. Second, the current empirical results demonstrate that hospital departments should invest in their capability to balance both the organization's efforts to pursue new knowledge and IT resources and their capability to take advantage of existing IT resources and assets. IT ambidextrous hospital departments are better equipped to identify, develop new innovative digital opportunities and patient services, and enhance patient agility. This development path is crucial for successful hospital departments that strive to enhance the patient's clinical journey and provide patients with fitting health services.

Finally, the results imply that hospital departments should actively manage digital technologies to strive for patient agility. The department's digital dynamic capability is crucial in the development of new digital patient service solutions. Hospital department managers should strive to be agile in the modern turbulent economic environment. Therefore, they should dedicate their resources to leverage this capability fully. This way, they are better equipped to search, identify, and absorb new technological innovations, integrate, process, and exchange patient information, use them for decision-making processes, and anticipate and respond fast to changes in the patient's health service needs. In essence, the digital dynamic capability is about developing the core competencies, knowledge, and skills to better process patient information, adequately respond to digital transformation, mastering the state-of-the-art digital technologies, and deliberately develop innovative patient services using digital technology.

## 6.3     Limitations and future work

Several limitations constrain this current work. These limitations could drive future research avenues. First, this study followed a cross-sectional design and used self-reported measures. Although this approach is similar to that of, e.g., (Faber, van Geenhuizen, & de Reuver, 2017; Wu, Chen, & Greenes, 2009), the collected data for all constructs from a single person could lead to self-reporting bias. Future work could embrace a matched-pair survey where different respondents and stakeholders fill in several parts of the survey. Triangulation of possible available data from public sources could also enrich the current insights and further strengthen and validate the empirical results.

Second, the present study did not investigate the impact of digital dynamic capability and patient agility on performance or patient service benefits, nor did it consider possible contingent attributes of organizational context in which the hospital departments. Future research may also wish to investigate these critical topics to unfold critical insights on leveraging these clinical practice capabilities. Also, comparing outcomes across various hospitals and healthcare organizations might further contribute to the study findings' generalizability. Future work could also involve the patient engagement and digital technology co-design perspectives (Donelan, DesRoches, Dittus, & Buerhaus, 2013; Egener et al., 2017; Papoutsi, Wherton, Shaw, Morrison, & Greenhalgh, 2020), as patient participation and engagement were currently out of scope.

Finally, the current study only gathered data from Dutch hospitals. Replication studies in other countries in Europe and other Non-Western regions could contribute to the generalization of the current outcomes.

## 6.4     Concluding remarks

Hopefully, these outcomes contribute to a better understanding of the relationship between the phenomenon of concurrently aiming for exploration (long-term perspective) and exploitation (current business environment perspective) in IT resource management and the mechanisms through which patient agility can be achieved in clinical practice. Scholars and medical practitioners can now benefit from these outcomes as patient sensing, and responding capabilities are crucial for hospitals to deliver high-quality patient value and streamlined patient journeys. This work is particularly relevant now, as hospitals worldwide need to transform healthcare delivery processes using digital innovations during the COVID-19 crisis.